\documentclass[aps,prd,twocolumn,showpacs,superscriptaddress,groupedaddress]{revtex4}  
\usepackage{graphicx}  
\usepackage{dcolumn}   
\usepackage{bm}        
\usepackage{amssymb}   
\usepackage{epsfig}
\usepackage{amsmath}

\newcommand{\be}{\begin{equation}}
\newcommand{\ee}{\end{equation}}
\newcommand{\beq}{\begin{eqnarray}}
\newcommand{\eeq}{\end{eqnarray}}

\begin{document}

\title{Testing a dissipative kinetic k-essence model}

\author{V\'ictor H. C\'ardenas}
\email{victor.cardenas@uv.cl}
\affiliation{ Instituto de F\'{\i}sica y Astronom\'ia, Universidad de Valpara\'iso,
Gran Breta\~na 1111, Valpara\'iso, Chile,}
\affiliation{Centro de Astrof\'isica de Valpara\'iso, Gran Breta\~na 1111, Playa Ancha,
Valpara\'{\i}so, Chile.}

\author{ Norman Cruz }
\email{norman.cruz@usach.cl}
\affiliation{Departamento de F\'{\i}sica, Universidad de Santiago de Chile,
Av. Ecuador 3493, 
Santiago, Chile}

\author{ J. R. Villanueva }
 \email{jose.villanuevalob@uv.cl}
\affiliation{ Instituto de F\'{\i}sica y Astronom\'ia, Universidad de Valpara\'iso,
Gran Breta\~na 1111, Valpara\'iso, Chile,}
\affiliation{Centro de Astrof\'isica de Valpara\'iso, Gran Breta\~na 1111, Playa Ancha,
Valpara\'{\i}so, Chile.}

\date{\today}

\begin{abstract}
In this work, we present a study of a purely kinetic k-essence
model, characterized basically by a parameter $\alpha$ in presence
of a bulk dissipative term, whose relationship between viscous
pressure $\Pi$ and energy density $\rho$ of the background follows a
polytropic type law $\Pi \propto \rho^{\lambda+1/2}$, where
$\lambda$, in principle, is a parameter without restrictions.
Analytical solutions for the energy density of the k-essence field
are found in two specific cases: $\lambda=1/2$ and
$\lambda=(1-\alpha)/2\alpha$, and then we show that these solutions
posses the same functional form than the non-viscous counterpart.
Finally, both approach are contrasted with observational data from
type Ia supernova, and the most recent Hubble parameter
measurements, and therefore, the best values for the parameters of
the theory are founds.

\end{abstract}



\maketitle


\section{Introduction}\label{intro}
At present, the scientific community dedicated to the study of the
universe have deep and intriguing questions unanswered. One of the
most fascinating corresponds to what we know as \textit{dark energy}
(DE) \cite{dereview,astier,martin,frieman,turner}, a component
designed to explain the current acceleration in the expansion of the
universe. In its simplest form, this can be described by a perfect
fluid with constant energy density, which leads to the useful
$\Lambda$ -- cold dark matter ($\Lambda$CDM) model, the simplest
model that fits a varied set of observational data. However, this
model has a high dependence to initial conditions that makes it
unnatural in many ways. For example, the current value for
$\Omega_{\Lambda}$ and $\Omega_DM$ are of the same order of
magnitude, a fact highly improbable, because the \textit{dark
matter} (DM) contribution decreases with $a^{-3}$, with $a(t)$ the
scale factor, meanwhile the cosmological constant remains constant.
This problem in particular is known as the cosmic coincidence
problem.
It is for this reason that many of the most sophisticated
experiments and instruments have been put in place; as the Dark
Energy Survey (DES) \cite{des}, the Baryon Oscillation Spectroscopic
Survey (BOSS) \cite{boss}, and the upcoming Large Synoptic Survey
Telescope (LSST) \cite{lsst} to mention some, all of them trying to
find new insights into the nature of dark energy.
In this context, the most natural way to understand the acceleration
of the universe, is to assume the existence of a dynamical
cosmological constant, or a theoretical model with a dynamical
equation of state parameter ($p/\rho=w(z)$). The source of this
dynamical dark energy could be both, a new field component filling
the universe, as a quintessence scalar field
\cite{quinta,ratra88,frieman95,turner97,caldwell,liddle1,steinhardt99},
or it can be produced by modifying gravity
\cite{modgrav,deffayet01,nojiri03,nojiri07,nojiri08,carroll04,song07}.
In this work, the so--called \textit{k-essence} model
\cite{armendariz99,deputter2007} is used, which is a type of
dynamical cosmological constant model, but where the source of its
dynamics comes from a non trivial kinetic term, as opposite to the
case of a typical quintessence model where the source is a different
scalar field potential, and then put it into the test with current
observational data from both, type Ia supernovae \cite{union2}, and
the most update Hubble parameter measurements \cite{farooqratra}.

Besides, if we focus, for example, in the dark sector as a whole, it
has been proved that the division of this sector into DM and DE is
merely conventional since exist a degeneracy between both
components, resulting from the fact that gravity only measures the
total energy tensor \cite{Kunz} (see also
\cite{Kunz1,kunz2,kunz3,kunz4,kunz5,kunz6,kunz7}). So, in the lack
of a well confirmed detection (nongravitational) of the DM only the
overall properties of the dark sector can be inferred from
cosmological data, at the background and perturbative level. This
results has driven the research to explore alternative models which
consider a single fluid that behaves both as DE and DM, the called
unified DM models (UDM). So this fluid must drive both the
accelerated expansion of the Universe at late times and the
formation of structures (see \cite{Bertacca} for a review of these
models).  Of course, a small speed of sound should be an essential
characteristic of a viable unified model in order to do not impede
the structure formation and to have a ISW effect signal compatible
with CMB observation
\cite{Bertacca1,Bertacca2,Bertacca3,Bertacca4,Bertacca5,Bertacca6}.

In this present work we will consider UDM models derived in the
framework of k-essence fields, common in effective field theories
arising from string theory and in particular in D-branes models
\cite{Callan,gibbons98,gibbons03,sen021,sen022}. This generalization
of the canonical scalar fields models can give rise to new dynamics
not possible in quintessence. In the context of cosmology, k-essence
was first studied as a model for inflation (k-inflation)
\cite{Picon}. K-essence  models has also addressed the problems of a
dynamical DE \cite{Picon1,chimento04} and the coincidence problem
\cite{Picon2,chiba}. For example, a particular case is the
Generalized Chaplygin gas (GCG) which appears as the simplest
tachyon field model, introducing in \cite{Chimento:2003ta}, with a
constant potential.  Moreover, k fields leads to a new Chaplygin
gases. Within the models investigated in order to unify DE and DM
are the GCG
\cite{Moschella01,bilic02,bento02,makler03,carturan03,amendola03,sandvik04,bertolami07}
and those known as purely kinetic models \cite{Scherrer,bertacca07}.
The unification of DE, DM and inflation has been addressed in
\cite{Bose,desantiago11}.

Another issue that emerges from the cosmological data is that the
exotic behavior of the universal fluid can be characterized by a
negative pressure and usually represented by the equation of state
$w= p/ \rho$, where $w$ lies very close to $-1$, most probably being
below $-1$. For example, the last Planck results give $w =
-1.13^{+0.13}_{ -0.10}$ and $w = -1.09±0.17$ $(95 \% CL)$ by using
CMB combined with BAO and Union2.1 data \cite{Susuki}, respectively,
for a constant $w$ model. In combination with SNLS3 data and $H_{0}$
measurement, the EoS for this dark component are $w =
-1.13^{+0.13}_{ -0.14}$ and $w = -1.24^{+0.18}_{-0.19}$ $(2\sigma
CL)$, respectively. The possibility of $w < -1$ is favored at the
$2\sigma$ level. These results are indicating that a phantom
behavior of the dark energy component can not ruled out from current
cosmological data.

As it was pointed out in~\cite{Cadwell} dark energy with a constant
EoS $w<-1$ leads to uncommon cosmological scenarios. First of all,
there is a violation of the dominant energy condition (DEC), since
$\rho+ p <0$. The energy density grows up to infinity in a finite
time, which leads to a big rip, characterized by a scale factor
blowing up in this finite time.  Nevertheless, sudden future
singularities are not necessarily produced by a fluids violating
DEC. Solutions which develop a big rip singularity at a finite time
without violate the strong-energy conditions $\rho>0$ and $\rho +3p
>0$ were found in~\cite{Barrow041,barrow042}. Studies of unified
dark matter models, which are generalizations of the Chaplygin gas,
presents EoS $w<-1$ but without a big rip type solution in
~\cite{Pedro}

Another mechanism that allows a violation of DEC is the existence of
dissipation within the cosmic fluids~\cite{Barrow87,barrow88}. In
the case of isotropic and homogeneous cosmologies, any dissipation
process in a FRW cosmology is scalar, and therefore may be modeled
as a bulk viscosity within a thermodynamical approach. The bulk
viscosity introduces dissipation by only redefining the effective
pressure, $p_{_{eff}}$, according to
$p_{_{eff}}= p+\Pi= p-3 \zeta H$,
where $\Pi$ is the bulk viscous pressure, $\zeta$ is the bulk
viscosity coefficient and $H$ is the Hubble parameter,
and $c = 8\pi\, G= 1$ (as in all the work). Since the
equation of energy balance is
$
\dot{\rho}+ 3 H (\rho+p+\Pi)=0$,
the violation of DEC, i.e., $\rho+p+\Pi<0$ implies an increasing
energy density of the fluid that fills the universe, for a positive
bulk viscosity coefficient. The condition $\zeta >0$ guaranties a
positive entropy production and, in consequence, no violation of the
second law of the thermodynamics~\cite{Pavon}.

Some investigations have considered that the viscous pressure can
drives the present acceleration of the Universe, so it can be used
to eliminate the dark energy component and to formulate unified dark
matter model with viscous pressure. In~\cite{Arturo09,Arturo10}, for
example, cosmological models where the only component is a
pressureless fluid with a variable and constant bulk viscosity was
confronted with the observational data. Nevertheless, the bulk
viscosity induces a large time variation of the gravitational
potential at late times which leads to inconsistencies with the
integrated Sachs-Wolfe (ISW) effect in such model
\cite{li09,Veltehn11,piattella11}.  In order to overcome this
problem, Velten \& Schwarz \citep{Velten} proposed a model with a
viscous cold dark matter and a cosmological constant, which acts
driving the accelerated expansion of the Universe. Our aim in this
work is to investigate UDM models derived in the framework of
k-essence fields which can also present dissipative effects.

Usually k-essence is defined as a quintessence, scalar field $\phi$
with a non-canonical kinetic energy associated with a Lagrangian
$\mathcal{L} = -V(\phi) F(X)$. In the subsequent calculations, we
shall restrict ourselves to the simple k-essence models for which
the potential $ V=V_{0}=$ constant. We also assume that $V_{0}=1$
without any loss of generality. One reason for studying k-essence is
that it is possible to construct a particularly interesting class of
such models in which the k-essence energy density tracks the
radiation energy density during the radiation-dominated era, but
then evolves toward a constant-density dark energy component during
the matter-dominated era. Such a behaviour can to a certain degree
solve the coincidence problem.

We investigate a dark energy model described by an effective
minimally coupled scalar field with a non-canonical kinetic term. If
for a moment we neglect the part of the Lagrangian containing
ordinary matter, the general action for a k-essence field $\phi$
minimally coupled to gravity is

\begin{equation}\label{actionKessence}
S= S_{G}+S_{\phi}=- \int d^{4}x \sqrt{-g}\,\left(\frac{R}{2}+
F(\phi,X)\right),
\end{equation}
where $F(\phi,X)$ is an arbitrary function of $\phi$ that
represents the k-essence action and $X=\frac{1}{2}\partial_{\mu}
\phi\partial^{\mu} \phi$ is the kinetic term.
We now restrict ourselves to the subclass of kinetic k-essence,
with an action independent of $\phi$
\begin{eqnarray}\label{actionKessence2}
S_{\phi}= -\int d^{4}x \sqrt{-g} \,F(X).
\end{eqnarray}
Unless otherwise stated, we consider
$\phi$ to be smooth on scales of interest so that $X = \frac{1}{2}
\dot{\phi}^2\geq0$. The energy-momentum tensor of the k-essence is
obtained by varying the action (\ref{actionKessence2}) with respect to the metric,
yielding
\begin{eqnarray}\label{energy-momentKessence}
T_{\mu\nu}=  F_{X}
\partial_{\mu}\phi\partial_{\nu}\phi - g_{\mu\nu}F,
\end{eqnarray}
where the subscript $X$ denotes differentiation with respect to
$X$. Identifying (\ref{energy-momentKessence}) as
the energy-momentum tensor of a perfect
fluid, we have the k-essence energy density $\rho$ and
pressure $p$

\begin{eqnarray}\label{RoKessence}
\rho&=& F-2XF_{X},\\\label{PressureKessence}
p&=&-F.
\end{eqnarray}
Throughout this paper, we will assume that the energy density is
positive so that $F-2 X F_X> 0$. The equation of state for the
k-essence fluid can be written as $p = w_{\phi}\rho=(\gamma_{\phi}-1)\rho$
with $F>0$,
\begin{eqnarray}\label{wKessence}
w_{\phi}=\gamma_{\phi}-1 = \frac{p}{\rho}= \frac{F}{2XF_{X}-F}.
\end{eqnarray}

\section{The k-essence model with dissipation}\label{sec:kke}

The Friedman--Lema\^itre--Robertson--Walker (FLRW) metric for an
homogeneous and isotropic flat universe is given by
\begin{eqnarray}\label{ndim}
ds^2=-dt^2+a(t)^2\left[dr^2+r^2 \left(d\theta^2 + sin^2 \theta\, d
\phi^2\right) \right],
\end{eqnarray}
where $a(t)$ is the scale factor and $t$ represents the cosmic time.
In the framework of the first order thermodynamic theory of
Eckart~\cite{Eckart} the field equations in the presence of bulk
viscous stresses yield
\begin{eqnarray}\label{tt}
\left(\frac{\dot{a}}{a}\right)^2=H^2= \frac{\rho}{3},
\end{eqnarray}
\begin{eqnarray}\label{rr}
\frac{\ddot{a}}{a}=\dot{H}+H^2=-\frac{1}{6} \left(\rho + 3
p_{_{eff}} \right),
\end{eqnarray}
where the effective pressure is given by
\begin{eqnarray}\label{Peff}
p_{_{eff}}=p+\Pi,
\end{eqnarray}
and
\begin{eqnarray}\label{Pi}
\Pi=-3 H \zeta,
\end{eqnarray}
is the bulk dissipative pressure and $\zeta$ the viscosity. In
what follows we will assume a power law dependence for the
viscosity in terms of the the density
\begin{equation}
\zeta = \zeta_0 \rho^{\lambda}, \label{zeta}
\end{equation}
where $\zeta_0$ is a positive semi-definite
constant with dimension $M^{1-\lambda}\,L^{3\lambda-1}\,T^{-1}$,
and $\lambda$ may take any  value.
For example, the most common values
are $\lambda=1/2$, i.e., $\zeta\propto \rho^{1/2}$ \citep{Li10,Brevik05,Brevik06,kuang,depaolis10} and $\lambda
=1$, i.e., $\zeta\propto \rho$ \citep{srd07,srd10}.
These values were chosen because leads to well known analytic solutions.
Therefore, the conservation
equation for the fluid
can be written as
\begin{eqnarray}\label{ConsEq}
\dot{\rho}+ 3 H (\rho+p+\Pi)=0.
\end{eqnarray}
In this work we consider the following function $F$ for the
k-essence field \citep{Chimento:2003ta}
\begin{equation}\label{fcn}
F(X)=\frac{1}{2\alpha-1}[X^{\alpha}-2 \alpha \alpha_0 \sqrt{X}],
\end{equation}
where $\alpha$ and $\alpha_0$ are two real constants. This generating
function exhibits a transition from a power law phase to a de Sitter stage,
inducing a modified Chaplygin gas. The explicit
equation of state can be obtained from Eqs. (\ref{RoKessence}) and
(\ref{PressureKessence})
\begin{equation}\label{EoSkessence}
p=(\mathfrak{n}-1)\rho-\mathfrak{n} \,\alpha_0\, \rho^{\frac{\mathfrak{n}-1}{\mathfrak{n}}},
\end{equation}
where the parameter $\mathfrak{n}$ is a function of the constant
$\alpha$, given by
\begin{equation}
\mathfrak{n}=\frac{2\alpha}{2\alpha-1}.
\label{pn}
\end{equation}
Obviously, the range of this parameter is
$1>\mathfrak{n}>0$, if $-\infty<\alpha<0$;
$0>\mathfrak{n}>-\infty$, if $0<\alpha<1/2$;
and $\infty>\mathfrak{n}>1$, if $1/2<\alpha<\infty$.

Of course, the speed of sound is affected by
the viscous pressure, which becomes
\begin{equation}\label{velson}
  v_{ef}^2=\frac{\partial p_{ef}}{\partial \rho}=
  v_{\phi}^2-(\lambda+1/2)\frac{\|\Pi\|}{\rho},
\end{equation}
where $v_{\phi}$ is the speed of sound in the purely
k-essence background \citep{Chimento:2003ta}, given by
\begin{equation}\label{sspk}
  v_{\phi}^2=(\mathfrak{n}-1)\left(1-\frac{\alpha_0}{\rho^{1/\mathfrak{n}}}\right).
\end{equation}

From Eqs.(\ref{tt} - \ref{fcn}),
together with the EoS (\ref{EoSkessence}),
we obtain the evolution equation
for $H$ in terms of the redshift,
\begin{equation}\label{hz}
  -a_0\,\frac{d H}{d x}+ a_1\, H +a_2 \, H^{\eta-1}+a_3\, H^{\beta-1}=0,
\end{equation}
where $x=\ln(1+z)$., and the coefficients are given by
\begin{equation}\label{coef}
  a_0=2,\, a_1=3\mathfrak{n},\, a_{2}=-3^{\frac{\mathfrak{n}-1}{\mathfrak{n}}}\alpha _{0}\mathfrak{n},\,
  a_{3}=-3^{\lambda +1}\zeta _{0},
\end{equation}
whereas the exponents reads
\begin{equation}\label{exp}
  \eta =2\left( \frac{\mathfrak{n}-1}{\mathfrak{n}}\right)=\frac{1}{\alpha},\qquad \beta =2\lambda +1.
\end{equation}

As a first observation, we note that there are two special values that yields
to well know equation without viscosity \citep{Chimento:2003ta}: $\beta=2$ ($\lambda=1/2$) and
$\beta=\eta$ ($\lambda=\frac{1-\alpha}{2\alpha}\neq 1/2$). These values leads to a single equation which posses
a generic structure for its quadrature given  by
\begin{equation}\label{eqg}
  \frac{dH}{dx}=A_1\,H+A_2 \,H^{\eta-1}\equiv A_1 \left( H+y\,H^{\eta-1}\right),
\end{equation}
where $y\equiv A_2/A_1$, and the new coefficients are given in terms
of the above by the following expressions
\begin{eqnarray}\label{A11}
  &&A_1=\frac{a_1+a_3}{a_0}=\frac{3}{2}(\mathfrak{n}-\sqrt{3}\,\zeta_0),\\ \label{A21}
  &&A_2=\frac{a_2}{a_0}=-\frac{3^{\frac{\mathfrak{n}-1}{\mathfrak{n}}}\alpha_0\,\mathfrak{n}}{2},\\ \label{y12}
  &&y=\frac{a_2}{a_1+a_3}=\frac{\alpha_0\,\mathfrak{n}}{3^{\frac{1}{\mathfrak{n}}}(\sqrt{3}\zeta_0-\mathfrak{n})},
\end{eqnarray}
for $\lambda=1/2$ (the model A), whereas
\begin{eqnarray}\label{A12}
  &&A_1=\frac{a_1}{a_0}=\frac{3}{1-2\lambda},\\ \label{A22}
  &&A_2=\frac{a_2+a_3}{a_0}=-\frac{3^{\lambda+\frac{1}{2}}}{2}\left(\frac{2\alpha_0}{1-2\lambda}+\sqrt{3}\zeta_0 \right),\\\label{y21}
  &&y=\frac{a_2+a_3}{a_1}=\frac{3^{\lambda-\frac{1}{2}}}{2}\, \left(\sqrt{3}\,\zeta_0 (2\lambda-1)-2\,\alpha_0 \right),
\end{eqnarray}
for $\lambda \neq 1/2$ (the model B). So,
a direct integration of Eq. (\ref{eqg}) leads to
\begin{equation}\label{hsc}
  H(z)=H_{0}\, \left[\frac{(1+z)^{\frac{2 A_1}{\mathfrak{n}}}-A_3}{1-A_3}\right]^{\frac{\mathfrak{n}}{2}},
\end{equation}
and therefore, the energy density is given by
\begin{equation}\label{dsc}
  \rho(z)=3\,H_{0}^2\, \left[\frac{(1+z)^{\frac{2 A_1}{\mathfrak{n}}}-A_3}{1-A_3}\right]^{\mathfrak{n}},
\end{equation}
where we have defined
\begin{equation}\label{a3}
A_3= \frac{\mathcal{R}}{1+\mathcal{R}},\qquad \left(\mathcal{R}=\frac{y}{H_{0}^{\frac{2}{\mathfrak{n}}}} \right).
\end{equation}

We note that the generic expression (\ref{dsc}) (or Eq. (\ref{hsc}))
has the form found by Chimento \cite{Chimento:2003ta}, and
obviously, these case is entirely recuperated by making $\zeta_0\rightarrow 0$
and $\lambda\rightarrow 0$.
A second observation
is that, in the case $\lambda \neq 1/2$ and by using
Eqs. (\ref{exp}) and (\ref{A12}),
the expression  (\ref{hsc})
takes the form
\begin{equation}\label{hsc2}
  H(z)=H_{0}\, \left[\frac{(1+z)^{3}-A_3}{1-A_3}\right]^{\frac{\mathfrak{1}}{1-2\lambda}}.
\end{equation}
Finally, there is a future singular value
of the redshift, say $z_s$, for which the Hubble function takes its
zero value:
\begin{equation}\label{zs}
  z_s=A_{3}^{\frac{\mathfrak{n}}{2A_1}}-1.
\end{equation}
We are restricted to the realistic values for
the future singularity, so we expect that  $-1<z_s<0$. Thus,
this condition impose that $y>0$, which implies that
$\mathfrak{n}<\sqrt{3}\zeta_0$ if $\lambda=1/2$, and
$\lambda>1/2+\alpha_0/(\sqrt{3}\zeta_0)$ if $\alpha=(2\lambda+1)^{-1}$.
In this context, notice that $\lambda=1$ (i. e. $\alpha=1/3$) leads to the condition
$2\,\alpha_0<\sqrt{3}\,\zeta_0$.

This kind of future singularity correspond to a novel type, because
although both the Hubble parameter (\ref{hsc}) and the energy
density (\ref{dsc} ) vanish at this redshift.

\section{Observational constraints}

In this section we use observational data to put some constraints in
the free parameters of the models. We use type Ia supernova data,
specifically the Union 2 data set \cite{union2}, and the most recent
Hubble parameter $H(z)$ measurements compiled in \cite{ratra},
consisting in 28 data points expanding a range in redshift
$0.015<z<2.3$.

The comoving distance from the observer to redshift $z$, in a flat
universe, is given by
\begin{equation}\label{comdistance}
r(z) =  \frac{c}{H_0} \int_0^z \frac{dz'}{E(z')},
\end{equation}
where $E(z)=H(z)/H_0$. The SNIa data give the luminosity distance
$d_L(z)=(1+z)r(z)$. Notice that the procedure we follow differ from
those used by Bandyopadhyay et al. \citep{Bandyopadhyay}. In this
work the authors define an intermediate parametrization for the
luminosity distance as a function of two parameters $\alpha, \beta$
which after the fitting is related to the physical parameters of the
model. Here we constrain directly the physical parameters of the
model.

We fit the SNIa with the cosmological model by minimizing the
$\chi^2$ value defined by
\begin{equation}
\chi_{SNIa}^2=\sum_{i=1}^{557}\frac{[\mu(z_i)-\mu_{obs}(z_i)]^2}{\sigma_{\mu
i}^2},
\end{equation}
where  $\mu(z)\equiv 5\log_{10}[d_L(z)/\texttt{Mpc}]+25$ is the
theoretical value of the distance modulus, and $\mu_{obs}$ is the
corresponding observed one.

From (\ref{dsc}) we can write down explicitly
\begin{equation}\label{ez}
E(z)=\left[\frac{(1+z)^{\frac{2 A_1}{\mathfrak{n}}}-A_3}{1-A_3}\right]^{\frac{\mathfrak{n}}{2}}.
\end{equation}
This form of the solution enable us to test both models at the same
time by reinterpreting the constants values. The best fit values using
both SNIa and $H(z)$ data
leads to a $\chi^2_{red}\simeq 0.96$,
and $A_1=1.50 \pm 0.15$,
$\eta=-0.29\pm 0.19$, and
$A_3=-2.9\pm 0.5$.

For the case $\lambda=1/2$, the free
parameters are three: $A_1$, the parameter
that changes with the model, $\eta$, which
is defined in (\ref{exp}), and $A_3$, defined
in (\ref{a3}). Straightforward calculations
lead to $\mathfrak{n}=0.87\pm 0.07$,
$\zeta_0=-0.075 \pm 0.069$, and
$\alpha_0=(2.96\pm 2.6)\times 10^{-4}$. Note that for the case $\lambda=1/2$ and since we have taken $G=1/8\pi$ and $c=1$, it is straightforward to see that parameter $\zeta_0$ is dimensionless.

Since the exponent in (\ref{ez}) reduces to $\frac{2
A_1}{\mathfrak{n}}=3$, in the case $\lambda \neq 1/2$ the free
parameters reduce to $A_3$ and $\eta$. For this reason, is not
possible to invert the equations completely, because this model is
described by three parameters, $\zeta_0, \alpha_0$ and $\lambda$. In
fact, from the  best fit, we can write down directly the value for
$\lambda=-0.65 \pm 0.08$. The other two parameters are tightly
related through the relation
\begin{equation}
\label{mod2}
\alpha_0=-\frac{y}{3^{\lambda-1/2}}+2\sqrt{3}(2\lambda-1)\zeta_0.
\end{equation}
Because the best value for parameter $y$ is large compared with the
second term in the right hand side (for reasonable positive values
of $\zeta_0$), the value for $\alpha_0$ is largely better
constrained than $\zeta_0$.

\begin{figure}[h]
\begin{center}
\includegraphics[width=75mm]{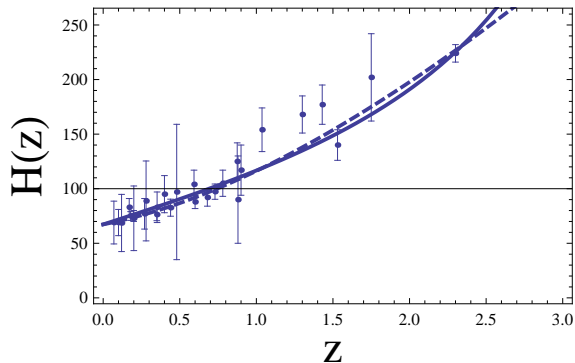}
\end{center}
\caption{Using the values of the best fit for each model, here we
display the theoretical curve of each model along the observational
data for $H(z)$. The continuous line is model A, and the dashed line
describe model B. It should be noted that the values of the best fit
was obtained using both measurements of $H(z)$ and supernovas. We
have adopted $h=0.673$ from the Planck Collaboration
\cite{Ade:2013zuv}.} \label{f1}
\end{figure}

\begin{figure}[h]
\begin{center}
\includegraphics[width=75mm]{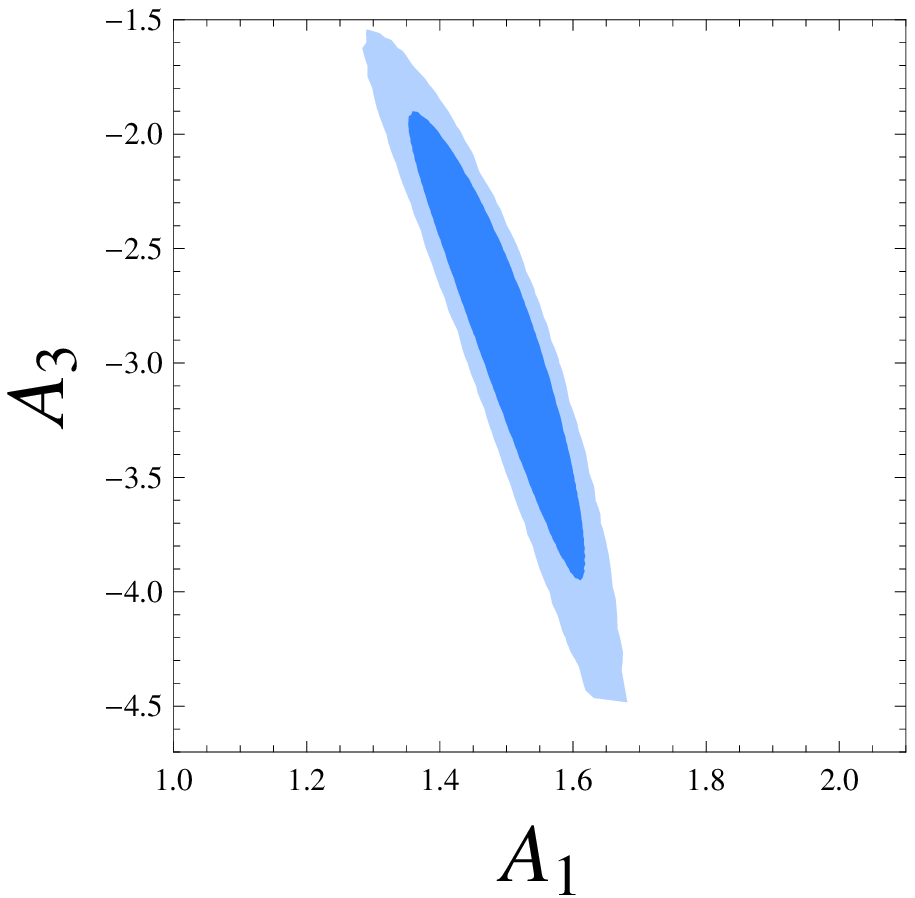}
\includegraphics[width=75mm]{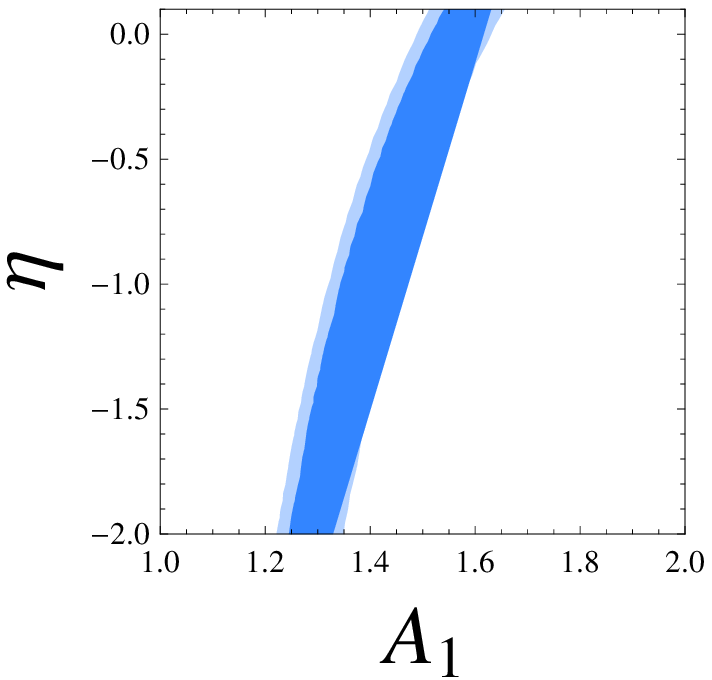}
\includegraphics[width=75mm]{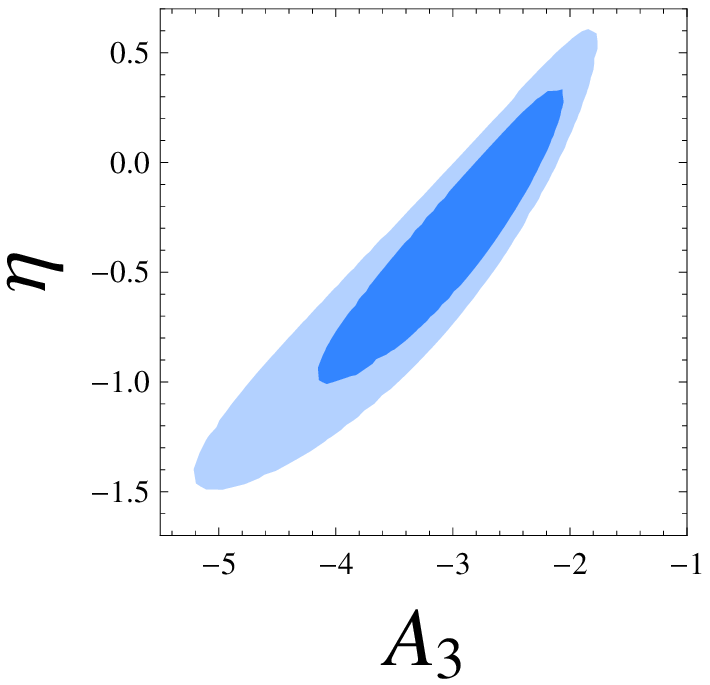}
\end{center}
\caption{Here we display the $68.27\% $ and $95.45\%$ confidence
regions for the parameters $A_1$, $A_3$ and $\eta$ for model A.}
\label{f2}
\end{figure}

\begin{figure}[h]
\begin{center}
\includegraphics[width=75mm]{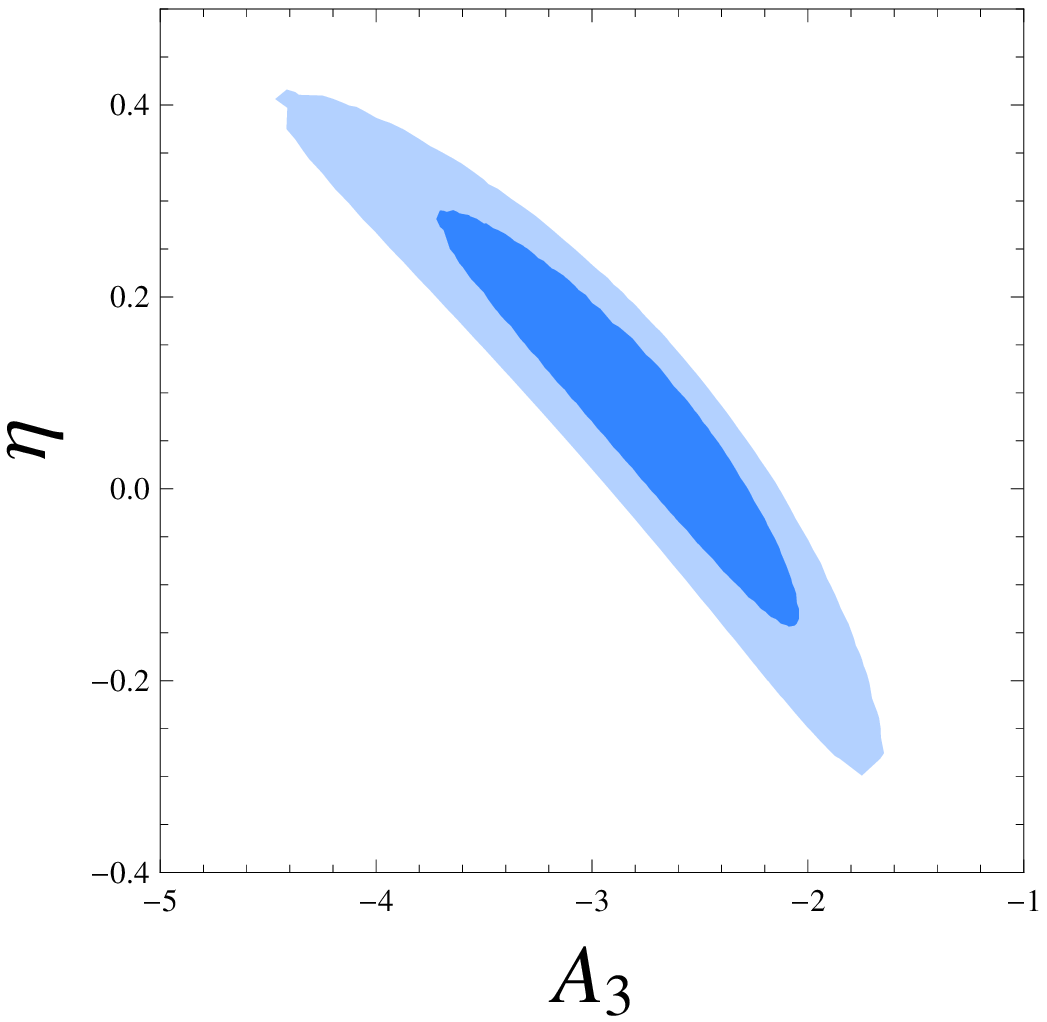}
\end{center}
\caption{Here we display the $68.27\% $ and $95.45\%$ confidence
regions for the parameters $A_3$ and $\eta$ for model B.} \label{f3}
\end{figure}

In order to make manifest the quality of the fit of our models, in
Figure (\ref{f1}) we show the theoretical curves of best fit for
each model together with the observational data of $H(z)$. There we
show the 28 data points measurements of the Hubble parameter
together with the best theoretical fit. We have to notice that
although the lines does not seems to follow the observational points
very well, this is because the best fit model was computed using
both SNIa data and $H(z)$ measurements, and the first data set
statistically weighs more than the second one, just because of the
number of data in each case. We also display in figure (\ref{f2})
the confidence level contours for the parameters $\eta$ and $A_3$ at
one and two $\sigma$, and in figure (\ref{f3}) the confidence
contours for the model B parameters.

In both cases, because the analysis was performed without imposing
external priors on the parameters, we found a preference for nearly
zero to negative values for the viscosity constant $\xi_0$.

Despite the strange results -- a negative value for the viscosity
constant -- after put in tension our solutions with the data, we
have confident that such a analysis can be done in the first place
for any other analytical solution that can be obtained in the
future. Of course, we do not expect to find that just our special
(analytical) solutions be the best fit to the data immediately.
Cosmology has entered into the era of precision cosmology, and with
it, the possibility to rule out effectively a particular
cosmological model.

\section{Conclusions}

We have analysed the general relations
for a model of k-essence generated by the function
$F(X)=\small{\frac{1}{2\alpha-1}}[X^{\alpha}-2 \alpha \alpha_0 \sqrt{X}]$
( proposed by Chimento \citep{Chimento:2003ta}),
when a dissipative pressure $\Pi\propto \rho^{\lambda+1/2}$ is included.
We found a family of analytical solutions in two special cases:
$\lambda=1/2$ and $\lambda=(1-\alpha)/2 \alpha$  (with $\alpha\neq 1/2$),
which coming from a similar differential equations and posses the same structure
that the non-viscous case (compare, for example,  Eq. (69) in reference \citep{Chimento:2003ta}
with  Eq. (\ref{dsc}) ).

Also, a quick observation of Eq. (\ref{velson})
shows that, depending on the value of $\lambda$,
the speed of sound may be greater ($\lambda<-1/2$), equal ($\lambda=-1/2$)
or less ($\lambda>-1/2$)
 than the speed
of sound without viscosity.
Obviously, a well behaved fluid requires $\lambda \geq -1/2$, which
corresponds to a consistency relation for $\lambda$.

As a light of observational data, we confront both analytical
solution with measurements of $H(z)$ and supernovas.
The best fit yields the following values for the parameters:
$\chi^2_{red}\simeq 0.96$,
$A_1=1.50 \pm 0.15$,
$\eta=-0.29\pm 0.19$, and
$A_3=-2.9\pm 0.5$. Therefore,
we obtain for the model A that
$\mathfrak{n}=0.87\pm 0.07$,
$\zeta_0=-0.075 \pm 0.069$, and
$\alpha_0=(2.96\pm 2.6)\times 10^{-4}$, while
for the model B we obtain that
$\lambda=-0.65 \pm 0.08$, and the other
two parameters are tightly related through
Eq. \ref{mod2}. So, both model present
a controversy with the physical meaning
($\zeta_0<0$ in the model A and $\lambda<-1/2$ in the model B).

This work can be improved in many ways. On one hand, we can attempt
an alternative way to obtain analytical solutions, a possibility we
are already studying using a novel technique proposed to solve
complex differential equations
\cite{iv1,iv2,iv3,iv4,iv5,RMT0}. In this case, we have the possibility to
consider $\lambda$ a free parameter, enhancing the parameter space
to find a best fit with data.

Certainly a more realistic model would also be interesting to study.
In this work we have considered a UDM model assuming nothing else
but a k-essence field is present. We can add explicitly a dark
matter term and/or a radiation component. We are interested in
testing if adding these terms would alleviate our concerns about the
sign of the viscosity coefficient.
\begin{acknowledgments}
N. C. acknowledges the hospitality of the Centre of Astrophysics
Valpara\'{\i}so (www.cav.uv.cl) and the Institute of Physics and
Astronomy of Universidad de Valpara\'{\i}so, where part of this work
was done.  We acknowledge the support to this research by Comisi\'on
Nacional de Investigaci\'on Cient\'ifica y Tecnol\'ogica through
FONDECYT grants Nos. 1110230 (VHC), 1140238 (NC), 11130695 (JRV),
and by DIUV project No 13/2009 (VHC).
\end{acknowledgments}


\begin{thebibliography}{}
\bibitem{dereview}

S.~Tsujikawa,
Dark Matter and Dark Energy: A Challenge for Modern Cosmology,
Astrophysics and Space Science Library Volume 370, 2011, pp 331-402
[arXiv: 1004.1493].


\bibitem{astier}
P. Astier and R. Pain,
Comptes Rendus Physique {\bf 13}, 521 (2012).

\bibitem{martin}
 J. Martin,
 Comptes Rendus Physique {\bf 13}, 566 (2012).

\bibitem{frieman}
J. Frieman, M. Turner and D. Huterer,
Ann. Rev. Astron. Astrophys. \textbf{46}, 385 (2008).

\bibitem{turner}
M.S. Turner and D. Huterer,
J. Phys. Soc. Jap. \textbf{76}, 111015
(2007).

\bibitem{des}
T.~Abbott {\it et al.}  [Dark Energy Survey Collaboration],
astro-ph/0510346.

\bibitem{boss}
  D.~Schlegel {\it et al.}  [with input from the SDSS-III Collaboration],
  arXiv:0902.4680 [astro-ph.CO].


\bibitem{lsst}
A.~Abate {\it et al.}  [LSST Dark Energy Science Collaboration],
[arXiv: 1211.0310 ].

\bibitem{quinta}
C. Wetterich,
Nucl. Phys. B \textbf{302}, 668 (1988).

\bibitem{ratra88}
B. Ratra and P. J. E. Peebles,
\prd {\bf 37}, 3406 (1988).

\bibitem{frieman95}
 J. A. Frieman, C. T. Hill, A. Stebbins and I. Waga,
\prl {\bf 75}, 2077 (1995).

\bibitem{turner97}
M. S. Turner and M. White,
\prd {\bf 56}, R4439 (1997).

\bibitem{caldwell}
 R. R. Caldwell, R. Dave and P. J. Steinhardt,
\prl {\bf 76}, 1582 (1998).

 \bibitem{liddle1}
A. R. Liddle and R. J. Scherrer,
\prd {\bf 59}, 023509 (1999).

\bibitem{steinhardt99}
P. J. Steinhardt, L. Wang, and I. Zlatev,
\prd {\bf 59}, 123504 (1999).



\bibitem{modgrav}
G. R. Dvali, G. Gabadadze, and M. Porrati,
Phys. Lett. B \textbf{485}, 208 (2000).

\bibitem{deffayet01}
C. Deffayet,
Phys. Lett. B\textbf{502}, 199 (2001).

\bibitem{nojiri03}
S. Nojiri  and S. D. Odintsov,
\prd \textbf{68}, 123512 (2003).

\bibitem{nojiri07}
S. Noriji and S.D. Odintsov,
Int. J. Geom. Meth. Mod. Phys. \textbf{4}, 115 (2007)

\bibitem{nojiri08}
S. Nojiri and S. D. Odintsov,
\prd \textbf{77}, 026007 (2008).


\bibitem{carroll04}
S. M. Carroll, V. Duvvuri, M. Trodden, and M.S. Turner,
\prd \textbf{70}, 043528 (2004).

\bibitem{song07}
Y. S. Song, W. Hu, and I. Sawicki,
\prd \textbf{75}, 044004 (2007).

\bibitem{armendariz99}
C. Armendariz-Picon, T. Damour and  T. Mukhanov,
Phys. Lett. B \textbf{458}, 209-218 (1999).


\bibitem{deputter2007}
R. de Putter and  E. V. Linder,
Astropart. Phys. \textbf{28}, 263-272 (2007) [arXiv: 0705.0400].

\bibitem{union2}
R. Amanullah {\it et al.},
\apj  {\bf 716}, 712 (2010)
  [arXiv: 1004.1711].

\bibitem{farooqratra}
O. Farooq and B. Ratra,
\apj  {\bf 766}, L7 (2013)
  [arXiv: 1301.5243].

\bibitem{Kunz}
M. Kunz,
\prd \textbf{71}, 023511 (2005).

\bibitem{Kunz1}
L. M. Reyes, J. E. Madriz Aguilar, L. A. Ure\~na-Lopez,
\prd \textbf{84}, 027503 (2011).

\bibitem{kunz2}
A. Aviles and J. L. Cervantes-Cota,
\prd \textbf{83}, 023510 (2011).

\bibitem{kunz3}
M. Kunz, A. R. Liddle, D. Parkinson, and C. Gao,
\prd \textbf{80}, 083533 (2009).

\bibitem{kunz4}
A. R. Liddle and L. A. Ure\~na-Lopez,
\prl \textbf{97}, 161301 (2006).

\bibitem{kunz5}
I. Wasserman,
\prd \textbf{66}, 123511 (2002).

\bibitem{kunz6}
C. Rubano and P. Scudellaro,
Gen. Relativ. Gravit. 34, 1931 (2002).

\bibitem{kunz7}
W. Hu and D. J. Eisenstein,
\prd \textbf{59}, 083509 (1999).

\bibitem{Bertacca}
D. Bertacca, N. Bartolo, and S. Matarrese,
Adv. Astron. 2010,  904379 (2010).

\bibitem{Bertacca1}
D. Bertacca and N. Bartolo,
J. Cosmol. Astropart. Phys.    \textbf{11}, 026 (2007).

\bibitem{Bertacca2}
H. Sandvik, M. Tegmark, M. Zaldarriaga and I. Waga,
\prd \textbf{69}, 123524 (2004).

\bibitem{Bertacca3}
R. J. Scherrer,
\prl \textbf{93}, 011301 (2004).

\bibitem{Bertacca4}
D. Pietrobon, A. Balbi, M. Bruni and C. Quercellini,
\prd \textbf{78}, 083510 (2008).


\bibitem{Bertacca5}
O. F. Piattella,
J. Cosmol. Astropart. Phys.   \textbf{03}, 012 (2010).

\bibitem{Bertacca6}
O. F. Piattella, D. Bertacca, M. Bruni, D. Pietrobon,
J. Cosmol. Astropart. Phys.   \textbf{01}, 014 (2010).


\bibitem{Callan}
J. Callan, Curtis G. and J. M. Maldacena,
Nucl. Phys. B \textbf{513}, 198-212 (1998).

\bibitem{gibbons98}
G. W. Gibbons,   Nucl. Phys. B \textbf{514} 603-639 (1998);

\bibitem{gibbons03}
G. W. Gibbons,
Rev. Mex. Fis. \textbf{49S1} 19-29 (2003).

\bibitem{sen021}
A. Sen,
J. High Energy Phys. \textbf{07}, 065 (2002).

\bibitem{sen022}
A. Sen,
 J. High Energy Phys. \textbf{04}, 048 (2002).

\bibitem{Picon} C. Armendariz-Picon, T. Damour and V. F. Mukhanov,
Phys. Lett. B \textbf{458}, 209-218 (1999).


\bibitem{Picon1}
C. Armendariz-Picon, V. Mukhanov, and P. J. Steinhardt,
\prl \textbf{85}, 4438-4441 (2000).

\bibitem{chimento04}
L. P. Chimento and  A.Feinstein,
Mod. Phys. Lett. A \textbf{19}, 761-768 (2004).

\bibitem{Picon2} C. Armendariz-Picon, V. F. Mukhanov and P. J. Steinhardt,
\prd \textbf{63}, 103510 (2001).

\bibitem{chiba}
T. Chiba, T. Okabe and M. Yamaguchi,
\prd \textbf{62}, 023511 (2000).


\bibitem{Chimento:2003ta}
L.~P.~Chimento,
\prd {\bf 69}, 123517 (2004).

\bibitem{Moschella01}
A. Y. Kamenshchik, U. Moschella and V. Pasquier,
Phys.Lett. B \textbf{511}, 265 (2001).

\bibitem{bilic02}
N. Bilic, G. B. Tupper and R. D. Viollier,
Phys. Lett. B \textbf{535}, 17 (2002).

\bibitem{bento02}
M. C. Bento, O. Bertolami and A. A. Sen,
\prd \textbf{66}, 043507 (2002).

\bibitem{makler03}
M. Makler, S. Quinet de Oliveira and I. Waga,
\prd \textbf{68}, 123521 (2003).

\bibitem{carturan03}
D. Carturan and F. Finelli,
\prd \textbf{68}, 103501 (2003).

\bibitem{amendola03}
L. Amendola, F. Finelli, C. Burigana and D. Carturan,
J. Cosmol. Astropart. Phys.   \textbf{07}, 005 (2003).

\bibitem{sandvik04}
H. Sandvik, M. Tegmark, M. Zaldarriaga and I. Waga,
\prd \textbf{69}, 123524 (2004).

\bibitem{bertolami07}
O. Bertolami , F. Gil Pedro and M. Le Delliou,
Phys. Lett. B \textbf{654}, 165-169 (2007).


\bibitem{Scherrer}  R. J. Scherrer,
\prl \textbf{93}, 011301  (2004).

\bibitem{bertacca07}
D. Bertacca and S. Matarrase, and M. Pietroni,
Mod. Phys. Lett. A \textbf{22}, 2893 (2007).

\bibitem{Bose}
N. Bose and A.S. Majumdar,
\prd \textbf{79},  103517 (2009).

\bibitem{desantiago11}
 J. De-Santiago  and J. L. Cervantes-Cota,
\prd \textbf{83}, 063502 (2011).

\bibitem{Susuki}
N. Suzuki et al.,
\apj \textbf{746}, 85 (2012).

\bibitem{Cadwell}
R. R. Cadwell, M. Kamionkowski and N. N. Weinberg,
\prl {\bf 91}, 071301 (2003).

\bibitem{Barrow041}
J. D. Barrow,
Class. Quantum Grav. {\bf 21}, L79  (2004).

\bibitem{barrow042}
 J. D Barrow,
 Class. Quantum Grav. {\bf 21}, 5619 (2004).

\bibitem{Pedro}
P. F. Gonz\'{a}lez-D\'{i}az,
Phys. Rev. {\bf 68}, 021303 (2003).

\bibitem{Barrow87}
J. D. Barrow,
Phys. Lett. B {\bf 180}, 335-339 (1987).

\bibitem{barrow88}
J. D Barrow,
Nucl. Phys. B {\bf 310}, 743 (1988).

\bibitem{Pavon}
W. Zimdahl and D. Pav\'on,
Phys. Rev. {\bf 61}, 108301 (2000).


\bibitem{Arturo09}
A. Avelino and U. Nucamendi,
J. Cosmol. Astropart. Phys.   \textbf{04}, 006 (2009).

\bibitem{Arturo10}
A. Avelino and U. Nucamendi,
J. Cosmol. Astropart. Phys.   \textbf{08}, 009 (2010).

\bibitem{li09}
B. Li and J. D. Barrow,
\prd \textbf{79}, 103521 (2009).

\bibitem{Veltehn11}
H. Velten and D. J. Schwarz,
J. Cosmol. Astropart. Phys.   \textbf{09}, 016 (2011).

\bibitem{piattella11}
O.F. Piattella. J.C. Fabris and W. Zimdahl,
J. Cosmol. Astropart. Phys.   \textbf{05}, 029 (2011).

\bibitem{Velten}
H. Velten and D. Schwarz,
\prd \textbf{86}, 08350 (2012).


\bibitem{Eckart}
C. Eckart,
Phys. Rev. {\bf 58}, 919 (1940).

\bibitem{Li10}
W. J. Li , Y. Ling, J. P. Wu and X. M. Kuang,
Phys. Lett.  B \textbf {687}, 1 (2010).

\bibitem{Brevik05}
I. Brevik and O. Gorbunova,
Gen. Rel. Grav. \textbf{37}, 2039 (2005).

\bibitem{Brevik06}
 I. Brevik,
Int. J. Mod. Phys. D \textbf{15}, 767 (2006).

\bibitem{depaolis10}
F. De Paolis, M. Jamil and A. Qadir,
Int. J. Theor. Phys. {\bf 49}, 621–632 (2010).

\bibitem{kuang}
X. M. Kuang and Y. Ling,
J. Cosmol. Astropart. Phys.   {\bf 10}, 024 (2009).

\bibitem{srd07}
S. del Campo, R. Herrera and D. Pav\'{o}n,
\prd\, \textbf{75}, 083518 (2007).

\bibitem{srd10}
S. del Campo, R. Herrera, D. Pav\'{o}n and J. R. Villanueva,
J. Cosmol. Astropart. Phys.\, \textbf{08}, 002 (2010).


\bibitem{ratra}
O. Farooq and B. Ratra:
\apj  {\bf 766}, L7 (2013).

\bibitem{Ade:2013zuv}
  P.~A.~R.~Ade {\it et al.}  [Planck Collaboration],
  arXiv:1303.5076 [astro-ph.CO].

\bibitem{Bandyopadhyay}
A. Bandyopadhyay, D. Gangopadhyay, A. Moulik,
Eur. Phys. J. C  {\bf 72}, 1943 (2012).


\bibitem{iv1}
B. Berndt,
Ramanujan's Notebooks, Part I.
Springer, New York (1985).


\bibitem{iv2}
T. Amdeberhan, O. Espinosa, I. Gonz\'{a}lez, H. Harrison,
V. H. Moll  and A. Straub,
The Ramanujan Journal  \textbf{29}, Issue 1-3 , pp 103-120 (2012).

\bibitem{iv3}
I. Gonz\'{a}lez, V. H. Moll and I. Schmidt,
A generalized Ramanujan Master Theorem applied to the evaluation of Feynman diagrams.
[arXiv: 1103.0588].



\bibitem{iv4}
I. Gonz\'{a}lez and V. H. Moll,
Adv. Appl. Math. {\bf45},  1, 50-73 (2010).

\bibitem{iv5}
I. Gonz\'alez I., V. Moll and A. Straub,
Contemporary Mathematics \textbf{517}, 157-171 (2010).

\bibitem{RMT0}
G. H. Hardy,
Ramanujan. Twelve Lectures on Subjects Suggested by His Life and Work.
3rd Ed. Chelsea, New York (1978).

\end{thebibliography}

\end{document}